\documentclass[final, 3p, times]{elsarticle}

\usepackage{amssymb}
\usepackage{mathtools}
\usepackage{mathrsfs}
\usepackage{physics}
\usepackage{braket}
\usepackage{subcaption}
\usepackage{hyperref}

\hypersetup{
    colorlinks=true,
}

\makeatletter
\newsavebox\myboxA
\newsavebox\myboxB
\newlength\mylenA

\newcommand*\xoverline[2][0.75]{%
    \sbox{\myboxA}{$\m@th#2$}%
    \setbox\myboxB\null
    \ht\myboxB=\ht\myboxA%
    \dp\myboxB=\dp\myboxA%
    \wd\myboxB=#1\wd\myboxA
    \sbox\myboxB{$\m@th\overline{\copy\myboxB}$}
    \setlength\mylenA{\the\wd\myboxA}
    \addtolength\mylenA{-\the\wd\myboxB}%
    \ifdim\wd\myboxB<\wd\myboxA%
       \rlap{\hskip 0.5\mylenA\usebox\myboxB}{\usebox\myboxA}%
    \else
        \hskip -0.5\mylenA\rlap{\usebox\myboxA}{\hskip 0.5\mylenA\usebox\myboxB}%
    \fi}
\makeatother

\biboptions{sort&compress}

\title{Non-perturbative flavor asymmetry in the nucleon and deuteron: \\ The light-front Hamiltonian effective field theory approach}

\author[1]{Xianghui Cao}
\author[5]{Shan Cheng}
\author[1]{Yihan Duan\corref{YihanAff}}
\author[1,7]{Yang Li\corref{speaker}}
\author[2,3,4]{Siqi Xu}
\author[2,3,6]{Xingbo Zhao}

\affiliation[1]{
organization={Department of Modern Physics, University of Science and Technology of China},
addressline={Hefei, Anhui},
postcode={230026},
country={China}
}

\affiliation[5]{organization={School of Physics and Electronics, Hunan University},
addressline={Changsha, Hunan},
postcode={410082},
country={China}}

\affiliation[7]{organization={Anhui Center for Fundamental Sciences in Theoretical Physics},
addressline={Hefei, Anhui},
postcode={230026},
country={China}}

\affiliation[2]{
organization={Institute of Modern Physics, Chinese Academy of Sciences},
addressline={Lanzhou},
postcode={730000},
country={China}
}

\affiliation[3]{
organization={School of Nuclear Science and Technology, University of Chinese Academy of Sciences},
addressline={Beijing},
postcode={100049},
country={China}
}

\affiliation[4]{
organization={Department of Physics and Astronomy, Iowa State University},
addressline={Ames, Iowa},
postcode={50011},
country={USA}
}

\affiliation[6]{organization={Advanced Energy Science and Technology Guangdong Laboratory},
addressline={Huizhou, Guangdong},
postcode={516000},
country={China}}

\cortext[speaker]{Speaker}
\cortext[YihanAff]{Present address: Department of Physics \& Astronomy, University of Pittsburgh, Pittsburgh, PA 15260, USA}

\begin{document}
\begin{frontmatter}

\begin{abstract}

We investigate non-perturbative multi-pion contributions to nucleon flavor asymmetry within the framework of Light-Front Hamiltonian Effective Field Theory (LFHEFT). Utilizing a Fock sector expansion, we systematically incorporate pionic degrees of freedom, with the nucleon-pion interactions governed by a scalar variant of chiral effective field theory. Our results demonstrate that the non-perturbatively calculated longitudinal momentum distributions exhibit significant deviations from leading-order perturbative predictions, emphasizing the importance of higher-order Fock components in describing the proton's sea quark structure. Furthermore, we demonstrate the feasibility of extending this framework to investigate nuclear effects in light nuclei, such as the deuteron. This unified approach provides a consistent basis for analyzing the interplay between intrinsic nucleon structure and nuclear modifications, potentially offering new insights into the flavor asymmetry observed in fixed-target and collider experiments.
\end{abstract}


\end{frontmatter}

\section{Introduction}

In the constituent quark model, the proton is traditionally described as a bound state of two up ($u$) quarks and one down ($d$) quark. With the advent of Quantum Chromodynamics (QCD), it became evident that the proton possesses a far more intricate internal structure. Beyond these valence quarks, the proton contains a sea of virtual gluons and quark-antiquark pairs.
Within the framework of perturbative QCD, these sea quarks are generated primarily through gluon splitting ($g \to q\bar q$). Because this mechanism is flavor-blind at leading order, it predicts a symmetric sea with an equal population of $\bar u$ and $\bar d$ antiquarks.
Therefore, any flavor asymmetry in the nucleon sea cannot be explained by perturbative evolution alone and non-perturbative QCD effects, such as pion cloud contributions or chiral symmetry breaking, are responsible \cite{Geesaman:2018ixo}.

Data from deep-inelastic scattering (DIS, \cite{NewMuon:1991hlj}) and Drell-Yan experiments ~\cite{NuSea:1998kqi,NuSea:2001idv, SeaQuest:2021zxb} provide compelling evidence for a $\bar{d}$ surplus. However, this conclusion has recently been challenged. Very recently, new measurements from $pp$ collisions at the Large Hadron Collider (LHC), which utilize the forward-backward asymmetry of the Drell-Yan process to decouple quark flavors, report no significant flavor asymmetry at similar kinematic regions~\cite{Ma:2025aga}.
This discrepancy has created a major tension in the field, challenging long-held assumptions about nucleon structure. One prominent explanation is that the previously observed asymmetry is not an intrinsic property of the proton, but rather an artifact of the effects of the nuclear environment within the deuterium targets utilized in fixed-target experiments. Alternatively, the tension may suggest a violation of isospin symmetry, which traditionally assumes that the $\bar{d}$ distribution in the neutron is identical to the $\bar{u}$ distribution in the proton. If this symmetry is broken within the deuteron environment, the extracted flavor ratios would be fundamentally shifted. In either scenario, the LHC results demand a radical re-evaluation of both our models of nucleon structure and our treatment of non-perturbative nuclear effects.

In this work, we introduce the light-front Hamiltonian effective field theory (LFHEFT) approach to investigate the parton structures within light nuclei, such as the deuteron. Our approach is based on the convolution formula, 
\begin{equation}
q(x) = \sum_h \int \frac{\dd z}{z} f_{h/H}(z/x) q_h^{(0)}(z) \equiv f_{h/H}\otimes q_h^{(0)},
\end{equation}
where $q_h^{(0)}$ represents quark PDFs in bare hadron and $f_{h/H}$ is the longitudinal momentum distribution (LMD) of $h$ within $H$, which can be accessed through the light-front wave functions (LFWFs), 
\begin{equation}
f_{h/H}(x) = \sum_n \int [\dd x_i \dd^2k_{i\perp}]_n \sum_{j\in h}\delta(x-x_j) \big|\psi_{n/H}(\{x_i, \vec{k}_{i\perp}\})\big|^2.
\end{equation}
These LFWFs are obtained non-perturbatively by solving the light-front Schrödinger equation, 
\begin{equation}\label{eqn:LF_schrodinger_equation}
    H_\text{LC} |\psi_H \rangle = M^2_H |\psi_H \rangle,
\end{equation}
where the light-cone Hamiltonian $H_\text{LC}$ is constructed using interactions derived from effective field theories (EFTs). 

The framework outlined above has been successfully implemented by Alberg and Miller within the pion cloud model to account for the flavor asymmetry observed in the SeaQuest experiments~\cite{Alberg:2021nmu}. Utilizing leading-order (LO) light-cone perturbation theory, the authors derived LFWFs from chiral EFT, where the physical proton is treated as a fluctuation into $|n\pi^+\rangle$ or $|\Delta^{++}\pi^-\rangle$ states. In this model, the flavor asymmetry arises primarily from the mass degeneracy breaking between the neutron and $\Delta$. 
Crucially, the pion cloud responsible for this asymmetry also mediates the nuclear binding within the deuteron. Consequently, a rigorous description of the asymmetry measured in SeaQuest must account for the longitudinal momentum distribution of nucleons within the nuclear environment. Furthermore, given the strength of the pion-nucleon coupling ($g_{\pi N} \approx 13$), it is natural to investigate the magnitude of multi-pion contributions to the flavor asymmetry. These interconnected issues -- nuclear binding and non-perturbative dynamics -- can be addressed consistently within the LFHEFT framework. In this paper, we present preliminary results investigating these effects.

\section{Non-perturbative nucleon sea} \label{sec:pion-cloud-model}

\begin{figure}
    \centering
    \begin{subfigure}[c]{0.32\textwidth}
        \includegraphics[width=\textwidth]{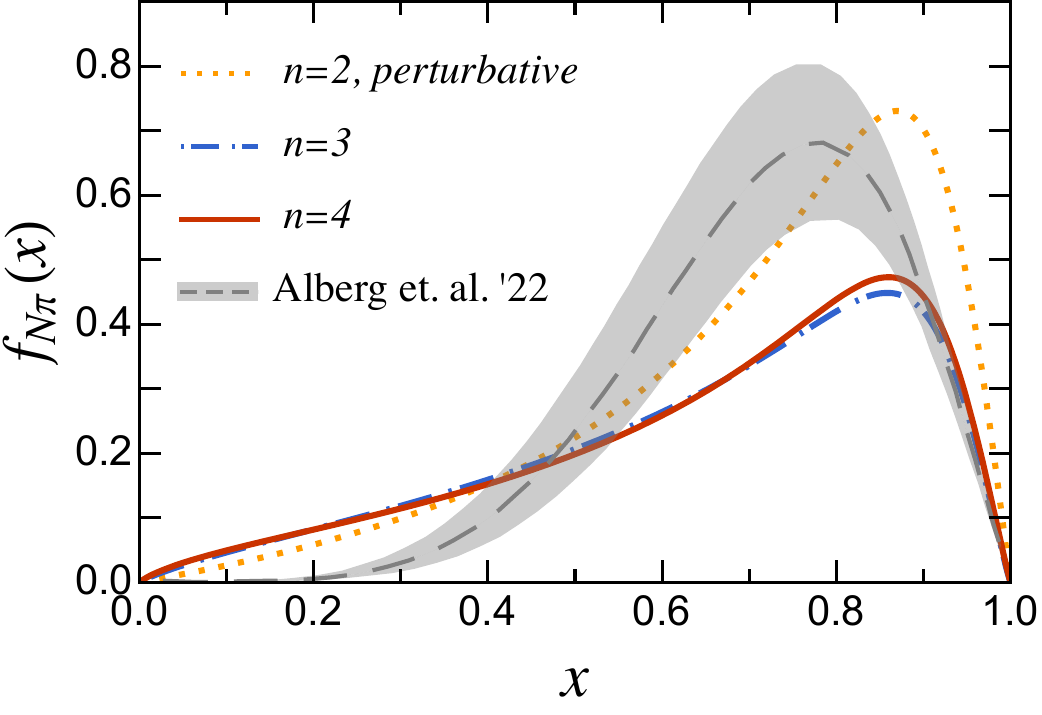}
    \end{subfigure}
    \begin{subfigure}[c]{0.32\textwidth}
        \includegraphics[width=\textwidth]{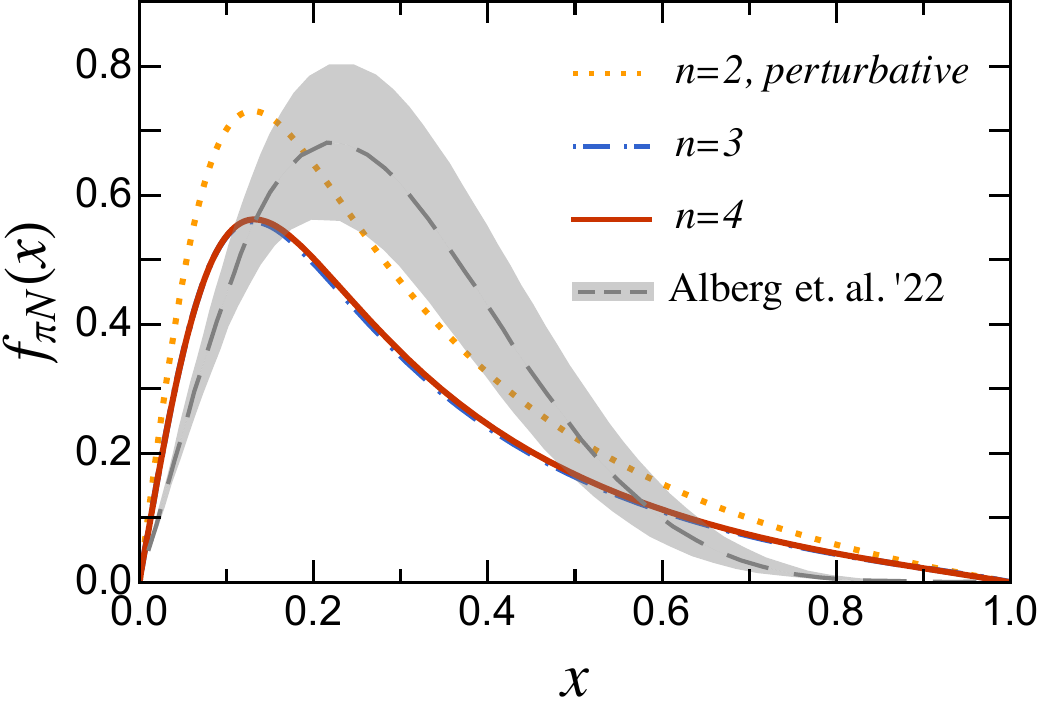}
    \end{subfigure}
    \begin{subfigure}[c]{0.32\textwidth}
        \includegraphics[width=\textwidth]{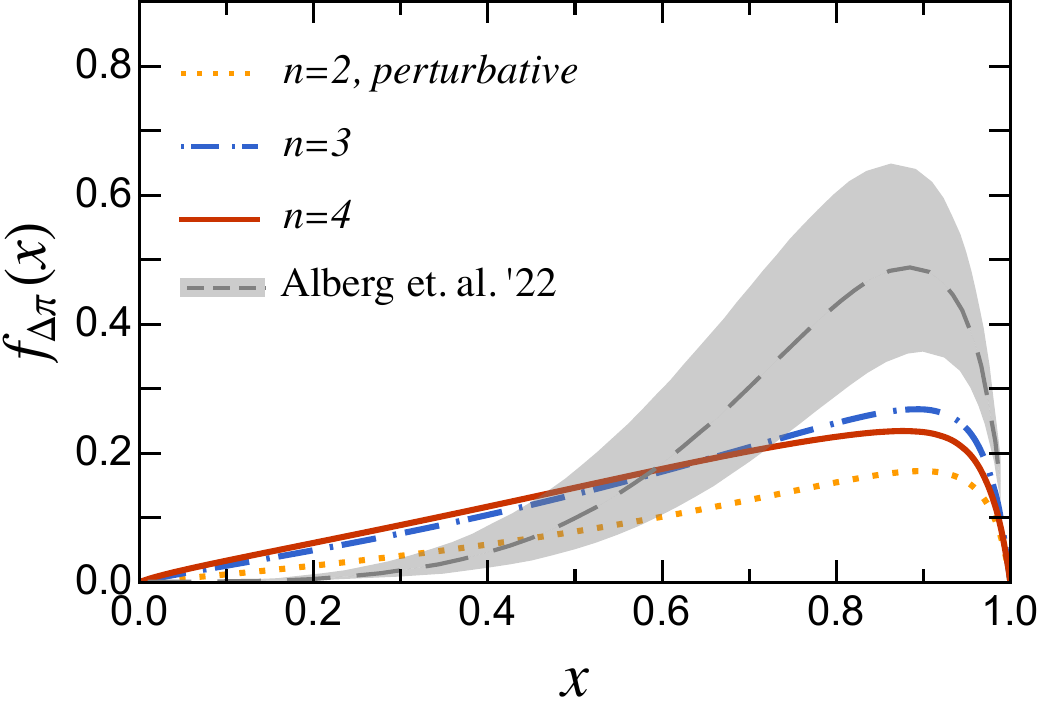}
    \end{subfigure}
    \caption{Longitudinal momentum distribution function of the nucleon ($f_{N\pi}$), pion ($f_{\pi N}$), and the $\Delta$ ($f_{\Delta \pi}$) inside the physical nucleon. Results are shown for two-body (baryon $+ \pi$), three-body (baryon $+ 2\pi$), and four-body (baryon $+ 3\pi$) truncations. The two-body truncation is equivalent to the leading-order light-front perturbation theory. For reference, we also include the leading-order perturbation theory results of Alberg \textit{et al}. obtained using a chiral Lagrangian~\cite{Alberg:2021nmu}.}
    \label{fig:LMDF}
\end{figure}

In the pion cloud model, the observed flavor asymmetry originates from non-nucleonic degrees of freedom -- specifically, intermediate pion-nucleon ($\pi N$) and pion-Delta ($\pi \Delta$) configurations within the physical nucleon~\cite{Alberg:2021nmu}. Within the LFHEFT framework, these contributions are systematically incorporated through a Fock sector expansion, which allows the nucleon to fluctuate into all dynamically permitted states. Consequently, the physical nucleon state can be represented as:
\begin{gather}
\ket{N}_{\text{ph}} = \ket{N} + \ket{N\pi} + \ket{N\pi\pi} + \ket{\Delta\pi} + \ket{\Delta\pi\pi} + \cdots,
\end{gather}
where the ellipsis denotes higher-order Fock sector components. The interactions between the nucleon, pion, and Delta resonance are governed by a scalar variant of chiral EFT:
\begin{gather}
\mathscr{L}_{\text{int}} = g_{N\pi} \xoverline{N} N \pi + g_{\Delta} \xoverline{N}\Delta \pi + g_{\Delta} \xoverline{\Delta} N \pi,\label{eq:Lagrangian}
\end{gather}%
In this formulation, $N$ and $\Delta$ are treated as complex scalar fields, while $\pi$ is represented by a real scalar field. The parameters $g_{N\pi}$ and $g_{\Delta}$ denote the dimensional coupling constants for the nucleon-pion and nucleon-Delta-pion vertices, respectively. For convenience, we define a dimensionless coupling constant:
$\alpha = \frac{g_{N\pi}^2}{16\pi M_N^2}$, where $M_N$ represents the nucleon mass.
The physical nucleon state is obtained by diagonalizing the light-cone Hamiltonian $H_\text{LC}$ Eq.~(\ref{eqn:LF_schrodinger_equation}).
To render the Hilbert space computationally tractable, the Fock space must be truncated to a finite number of sectors. The appropriate level of truncation is determined empirically by monitoring the convergence of the computed observables. As demonstrated in Ref.~\cite{Li:2015iaw}, this equation was solved using up to a four-body truncation ($\ket{N} + \ket{N\pi} + \ket{N\pi\pi} + \ket{N\pi\pi\pi}$), with robust convergence already evident at the three-body level ($\ket{N} + \ket{N\pi} + \ket{N\pi\pi}$). Consequently, we adopt the four-body truncation as the basis for the non-perturbative results presented in this work.

The quark distribution function within the physical nucleon is expressed via the convolution formula:
\begin{equation}
q_N(x) = Z q_N^{(0)}(x) + \sum_B \int_x^1 \frac{\mathrm{d}y}{y} \left[ f_{\pi B}(y) q_\pi(x/y) + f_{B\pi}(y) q_B(x/y) \right],
\end{equation}
where $Z$ represents the renormalization constant of the bare one-nucleon Fock component $\ket{N}$, and $q_N^{(0)}(x)$ is the bare nucleon's quark distribution. The terms $q_{\pi}(x)$ and $q_B(x)$ denote the quark distributions within the pion and baryon ($B = N, \Delta$), respectively, while $f_{\pi B}$ and $f_{B\pi}$ represent the corresponding pionic and baryonic LMDs. 
The LMD for a specific hadron can be computed directly from the many-body LFWFs:
\begin{equation}
f(x) = \sum_n \frac{1}{(n-1)!} \prod_{i=1}^n \int \frac{\mathrm{d} x_i \mathrm{d}^2 k_{i\perp}}{(2\pi)^3 2x_i} 2\delta(x-x_n) (2\pi)^3 \delta\left(\sum_i x_i - 1\right) \delta^{(2)}\left(\sum_i \vec{k}_{i\perp}\right) |\psi_n({x_i, \vec{k}_{i\perp}})|^2,
\end{equation}
where $\psi_n(\{x_i, \vec{k}_{i\perp}\})$ denotes the $n$-body LFWF, and the $n$-th particle is identified as the constituent of interest. Here, we utilize the standard light-front variables $x^{\pm} = x^0 \pm x^3$ and $\vec{x}_{\perp} = (x^1, x^2)$, defining $\vec{k}_{i\perp} = \vec{p}_{i\perp} - x_i \vec{P}_\perp$ as the intrinsic transverse momentum of the $i$-th constituent.

\begin{figure}
    \centering
    \begin{subfigure}[c]{0.32\textwidth}
        \hspace{-0.08\textwidth}
        \includegraphics[width=\textwidth]{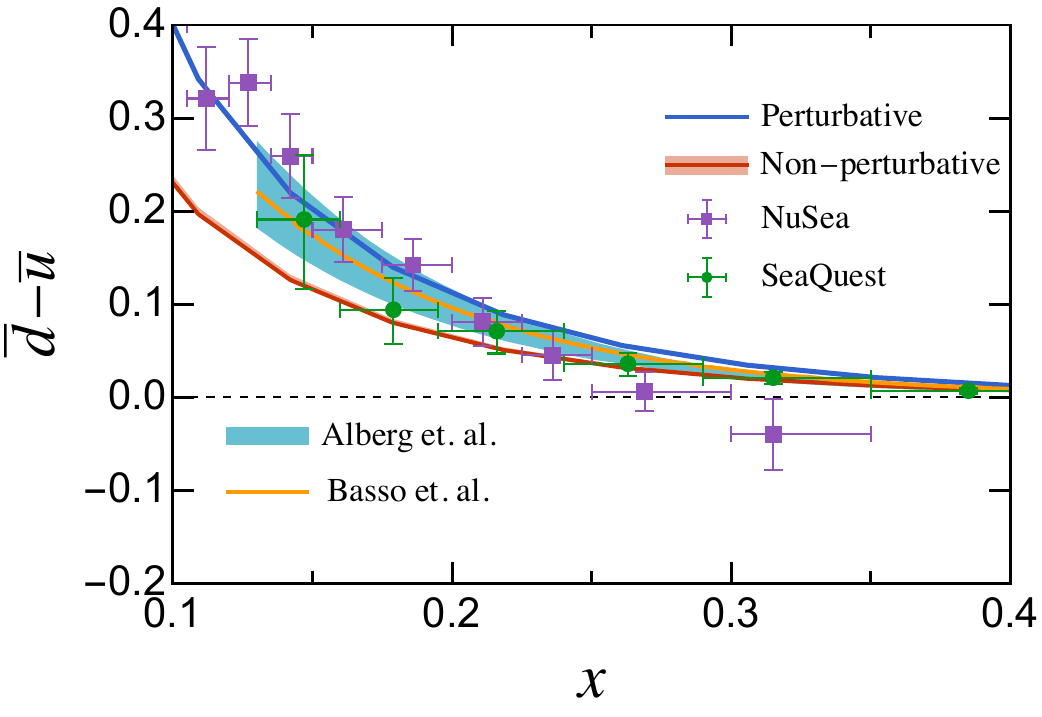}
    \end{subfigure}
    \hspace{0.08\textwidth}
    \begin{subfigure}[c]{0.32\textwidth}
        \hspace{-0.08\textwidth}
        \includegraphics[width=\textwidth]{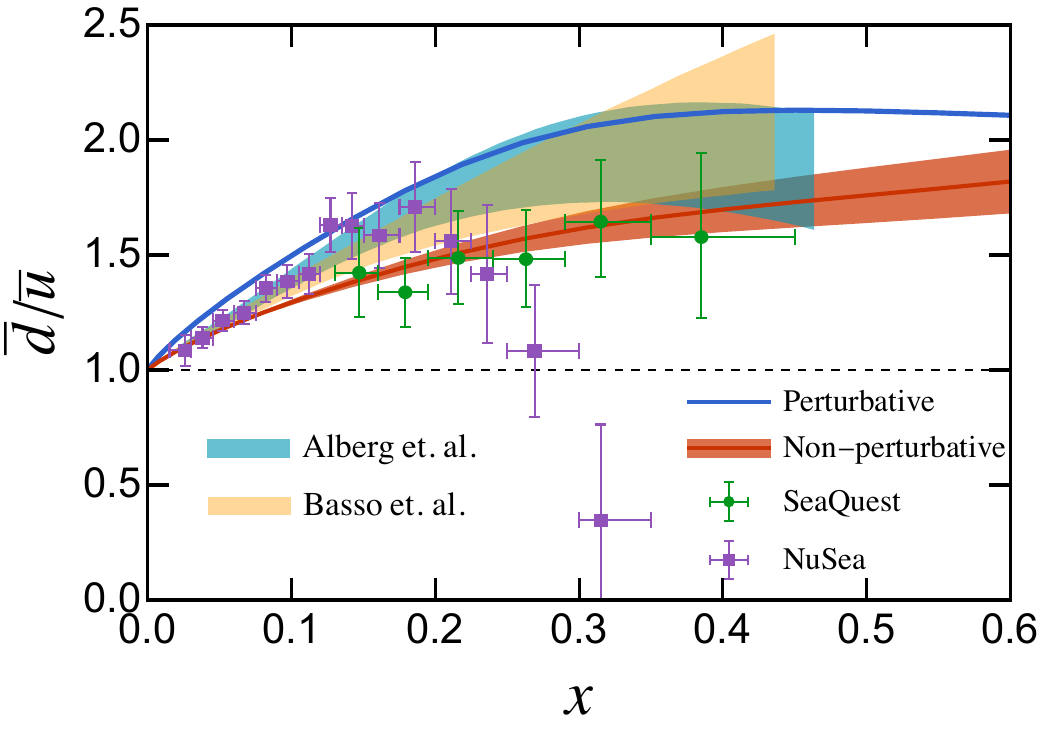}
    \end{subfigure}
    \caption{The distributions $\bar{d}(x) - \bar{u}(x)$ and $\bar{d}(x)/\bar{u}(x)$, calculated from the perturbative (two-body) and non-perturbative (four-body) solutions of our scalar model, compared with the NuSea~\cite{NuSea:1998kqi} and SeaQuest~\cite{SeaQuest:2021zxb} experimental data at the scale $Q^2 = 25.5~\text{GeV}^2$. The band for the non-perturbative results reflects the difference between the three-body and four-body truncations, illustrating the good convergence of the Fock sector expansion. Results from the leading-order light-front perturbation theory of Alberg \textit{et al}.~\cite{Alberg:2021nmu} and from the statistical model of Basso \textit{et al}.~\cite{Basso:2015lua} are also shown for comparison.}
    \label{fig:aq-asymmetry}
\end{figure}

In Fig.~\ref{fig:LMDF}, we present the LMDs for the nucleon, pion, and $\Delta$ resonance across two-, three-, and four-body truncations. The close agreement between the three-body and four-body results indicates a favorable convergence trend. Consequently, we adopt the four-body truncation as our primary non-perturbative result and estimate the theoretical uncertainty from the residual difference between the three-body and four-body calculations. The non-perturbative couplings are set to $g_{N\pi} = 6.0$ GeV ($\alpha = 0.8$) and $g_\Delta = 6.5$ GeV to reproduce the SeaQuest $\bar{d} - \bar{u}$ data~\cite{SeaQuest:2021zxb}. For the quark PDFs, we follow the methodology of Alberg et al., utilizing the pion and bare proton PDFs at a scale of $\mu^2 = 25.5$ GeV$^2$~\cite{Alberg:2021nmu}.
The two-body truncation corresponds to leading-order light-front perturbation theory. As illustrated in Fig.~\ref{fig:LMDF}, the non-perturbative LMDs deviate significantly from their perturbative counterparts, underscoring the essential role of multi-pion components. Although the perturbative LMDs in our scalar  model differ from the $\chi$EFT results of Alberg et al., both the perturbative and non-perturbative results in our framework successfully describe the SeaQuest data, as shown in Fig.~\ref{fig:aq-asymmetry}. The integrated flavor asymmetry over the kinematic range $0.13 < x < 0.45$ is found to be $\int_{0.13}^{0.45}\mathrm{d}x [\bar{d}(x) - \bar{u}(x)] = 0.0122(7)$, which is consistent with the SeaQuest measurement of $0.0159(60)$ reported in Ref.~\cite{FNALE906:2022xdu}. 

While the $\bar{d} - \bar{u}$ distribution shows that perturbative and non-perturbative predictions are nearly degenerate and align well with both NuSea and SeaQuest data, the $\bar{d}/\bar{u}$ ratio reveals moderate discrepancies, particularly in the high-$x$ region. Notably, the difference between the LO perturbative results and the three-body truncation is substantially larger than the difference between the three-body and four-body truncations. This suggests that the flavor asymmetry is highly sensitive to non-perturbative multi-pion components. The inclusion of $B + 2\pi$ configurations therefore exerts a significant impact on the predicted sea quark distributions.

\section{Nuclear effects within the deuteron} \label{sec:deuteron-wave-function}

Following the same framework used in the physical nucleon sector, we construct the deuteron state vector from the Fock expansion:
\begin{equation}
    |D\rangle = |NN\rangle + |NN\pi\rangle + |NN\pi\pi\rangle + \cdots 
\end{equation}
Solving the light-front Schrödinger equation of the deuteron as a strongly coupled few-body problem is highly non-trivial. As a preliminary step, we restrict the expansion to the first two Fock components: $|NN\rangle + |NN\pi\rangle$.
Rather than solving these coupled equations directly, we utilize the Wilson-Bloch method to construct an effective Hamiltonian acting within the reduced two-body subspace via a similarity transformation~\cite{bloch1958theorie,Chakrabarti:2001yh}. This methodology offers two significant advantages: The resulting effective Hamiltonian is Hermitian. The energy denominators are decoupled from the unknown eigenvalue, simplifying the numerical implementation~\cite{Chakrabarti:2001yh}. By integrating out the three-body ($|NN\pi\rangle$) sector, the effective description is reduced entirely to nucleonic degrees of freedom.
The resulting eigenvalue equation, incorporating the effective interaction kernel, is expressed as:
\begin{multline}
    \left[ \frac{\vec{k}_\perp^2 + m_N^2 + \sum_r(m_N^2 - x(s - M^2_D))}{x} + \frac{\vec{k}_\perp^2 + m_N^2 + \sum_r(m_N^2 - (1-x)(s - M_D^2))}{1-x} \right] \psi_D(x, \vec{k}_\perp)\\
    - \frac{4\alpha m_N^2}{\pi Z_2} \int_0^1 \frac{\dd x^\prime}{2x^\prime(1-x^\prime)} \int_0^\infty k_\perp^\prime \dd k_\perp^\prime K_\sigma(x, k_\perp; x^\prime, k_\perp^\prime) \psi_D(x^\prime, k_\perp^\prime) = M_D^2 \psi_D(x, k_\perp).
    \label{eq:deutron-EOF-Bloch}
\end{multline}
Here, $M_D$ denotes the deuteron mass and $m_N$ represents the nucleon mass. The terms $\Sigma_r(p^2)$ and $Z_2$ correspond to the self-energy and the field renormalization constant, respectively; their explicit analytical expressions are provided in the appendix of Ref.~\cite{Karmanov:2016yzu}. The effective interaction kernel, $K_\sigma$, is defined as:
\begin{gather}
    K_\sigma(x, k_\perp; x^\prime, k_\perp^\prime) = \frac{1}{\sqrt{A^2 - B^2}} \left( \frac{B}{A + \sqrt{A^2 - B^2}} \right)^{|\sigma|},
\end{gather}
where $\sigma$ is the magnetic projection of the deuteron. The scalar functions $A$ and $B$ are,
\begin{gather}
    A = \frac{1}{2}(x + x^\prime - 2xx^\prime) \left( \frac{k_\perp^2}{x(1-x)} + \frac{k_\perp^{\prime 2}}{x^\prime(1-x^\prime)} \right) + \frac{1}{2} m^2 (x - x^\prime)^2 \left( \frac{1}{xx^\prime} + \frac{1}{(1-x)(1-x^\prime)} \right),\quad 
    B = 2k_\perp k_\perp^\prime.
\end{gather}

\begin{figure}
    \centering
    \begin{subfigure}[c]{0.32\textwidth}
        \hspace{-0.08\textwidth}
        \includegraphics[width=\textwidth]{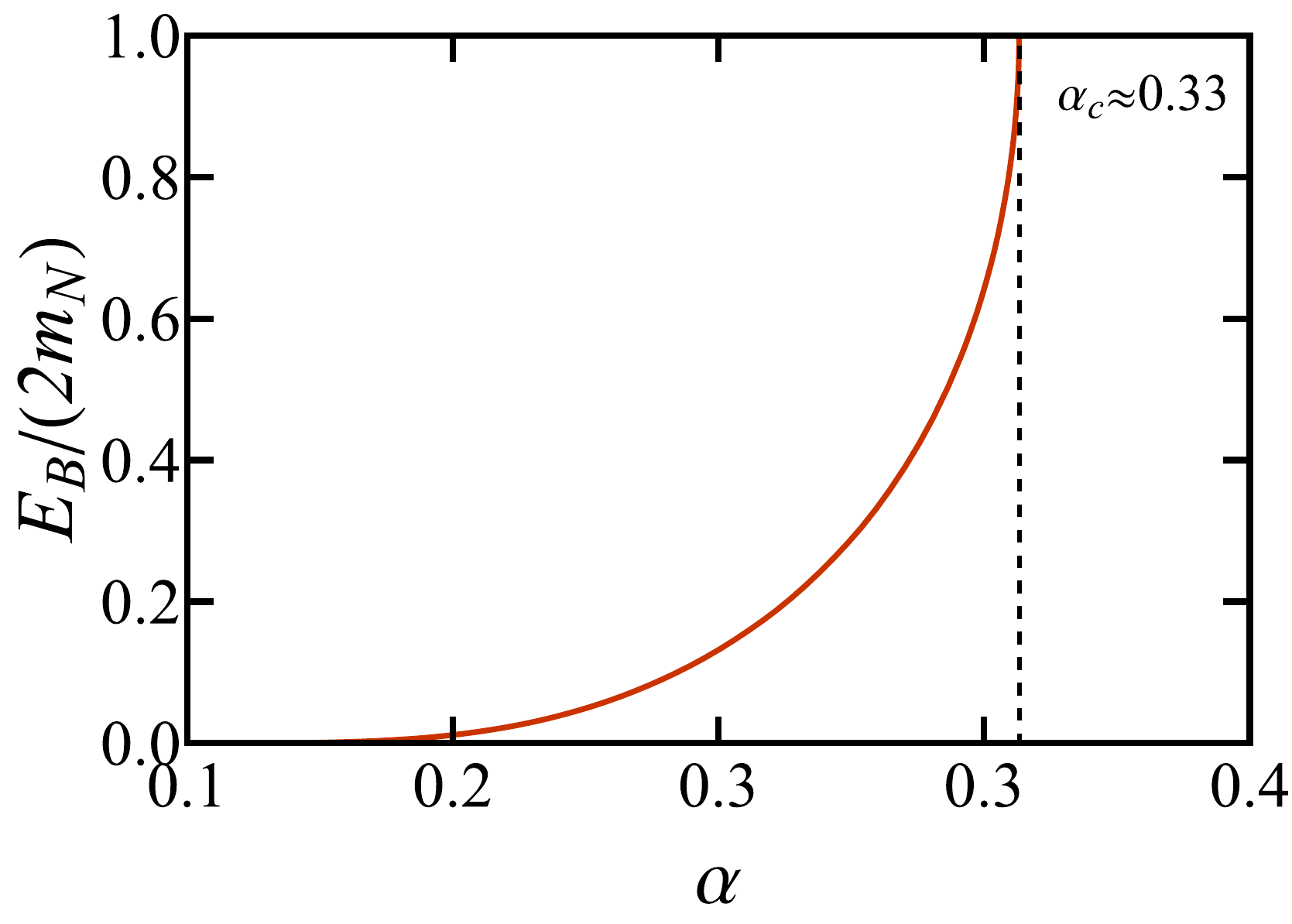}
        \subcaption{}
        \label{fig:alpha_vs_M}
    \end{subfigure}
    \hfill
    \begin{subfigure}[c]{0.32\textwidth}
        \hspace{-0.08\textwidth}
        \includegraphics[width=\textwidth]{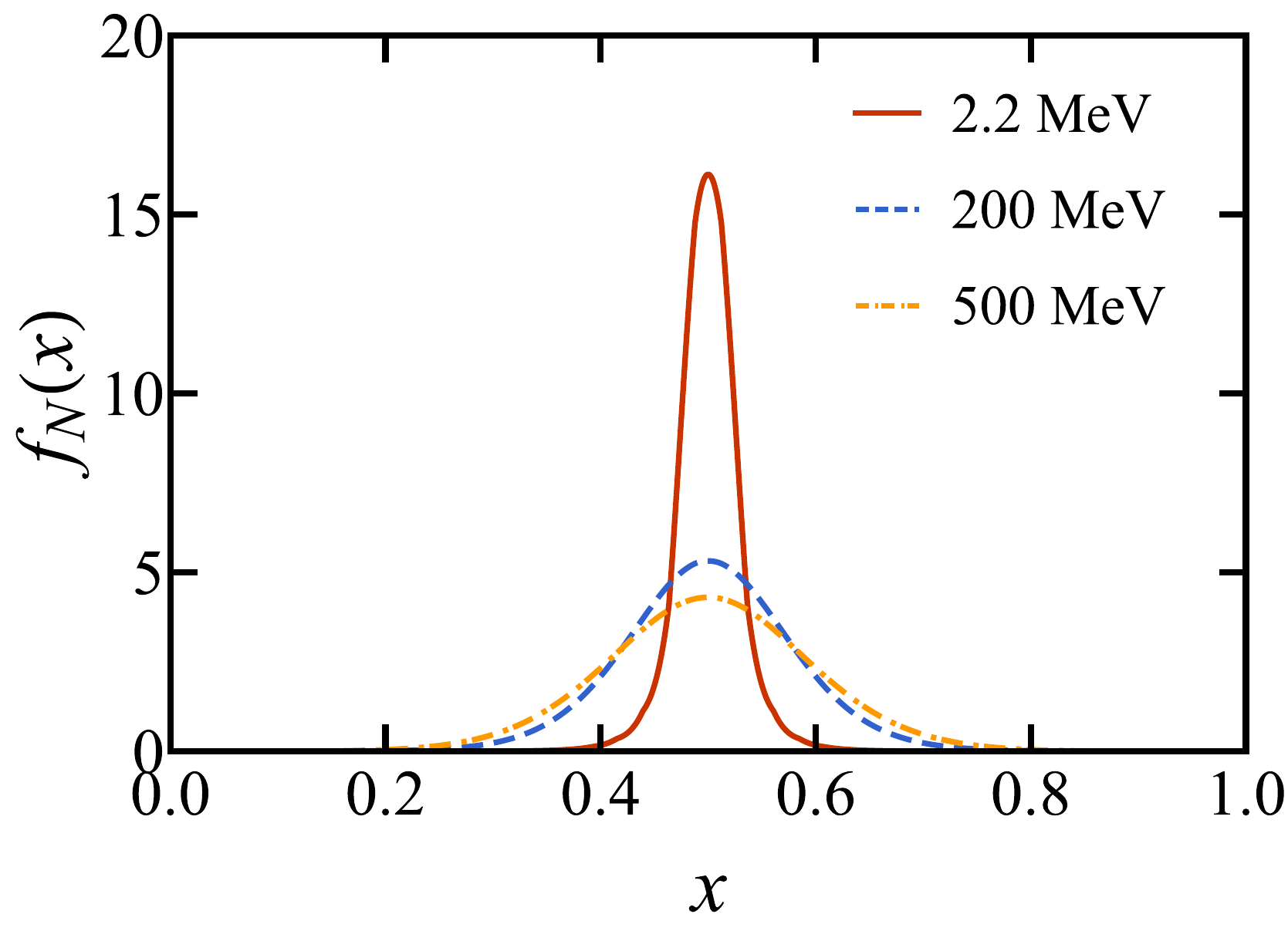}
        \subcaption{}
        \label{fig:deuteron-fn}
    \end{subfigure}
    \hfill
    \begin{subfigure}[c]{0.32\textwidth}
        \includegraphics[width=\textwidth]{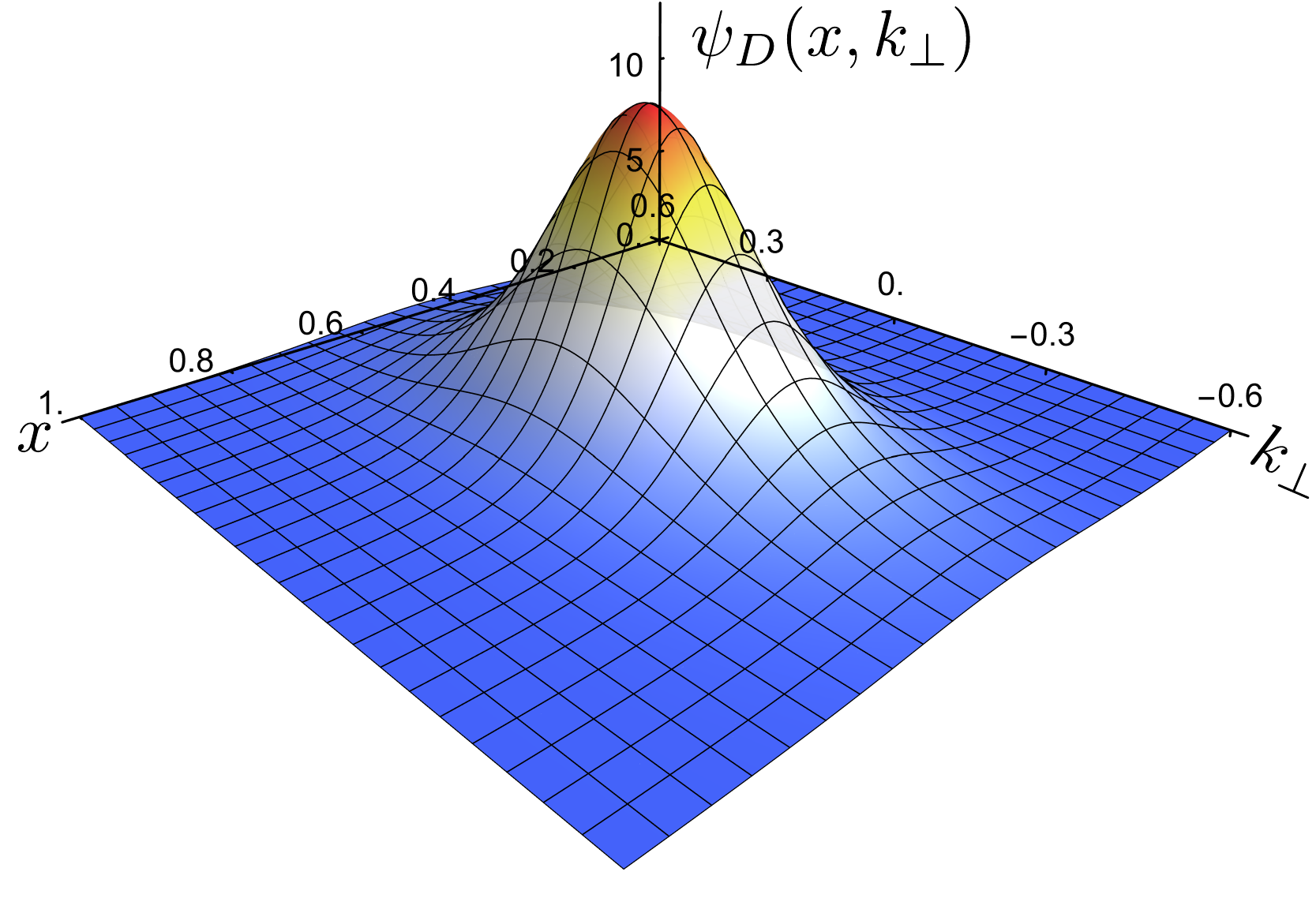}
        \subcaption{}
        \label{fig:deuteron-LFWF-EB200}
    \end{subfigure}
    \caption{(\subref{fig:alpha_vs_M}) Binding energy of the scalar deuteron as a function of the coupling constant $\alpha$. Beyond the critical coupling $\alpha_c \approx 0.33$, the system becomes unstable. (\subref{fig:deuteron-fn}) Longitudinal momentum distribution function of the scalar nucleon inside the scalar deuteron for $E_B = 2.2~\text{MeV}$ ($\alpha \approx 0.14$), $E_B = 200~\text{MeV}$ ($\alpha \approx 0.24$) and $E_B = 500~\text{MeV}$ ($\alpha \approx 0.28$). (\subref{fig:deuteron-LFWF-EB200}) The deuteron light-front wave function obtained from the Bloch effective Hamiltonian. The binding energy is $E_B \approx 200~\text{MeV}$ $(\alpha \approx 0.24)$.}
\end{figure}

Equation~\eqref{eq:deutron-EOF-Bloch} was solved for coupling constants in the range $0.12 < \alpha < 0.33$; outside this interval, the interaction is either too weak or too strong to support a stable bound state. The relationship between the ground-state binding energy, $E_B = 2m_N - M_D$, and the coupling constant $\alpha$ is depicted in Fig.~\ref{fig:alpha_vs_M}. To reproduce the physical deuteron binding energy of $E_B^{(D)} \approx 2.2$ MeV, the required coupling constant is approximately $\alpha \approx 0.14$. However, within a three-body truncation, different theoretical treatments can result in significant discrepancies in the predicted binding energy~\cite{Bakker:2002ih}. Notably, a recent AdS/QCD model suggests that a binding energy of approximately $200$ MeV may be necessary to reconcile theoretical predictions with experimental deuteron data~\cite{Kaur:2025css}. To investigate how a larger binding energy affects the nuclear structure, we computed the LMD of the scalar nucleon within the scalar deuteron for binding energies of $2.2$ MeV, $200$ MeV, and $500$ MeV. As shown in Fig.~\ref{fig:deuteron-fn}, the LMD at $2.2$ MeV is highly localized, closely resembling the non-relativistic limit. Increasing the binding energy to $200$ MeV broadens the distribution significantly, thereby exerting a substantial influence on the quark distributions within the deuteron. A further increase to $500$ MeV, however, produces only a marginal additional effect. These findings suggest that a single-pion-exchange mechanism alone may be insufficient to explain the observed flavor asymmetry at the physical deuteron binding energy. As higher Fock sectors -- representing multi-pion exchanges -- are included, the system is expected to become more tightly bound, leading to a broader LMD. This observation underscores the necessity of incorporating higher-order Fock components in future analyses. Finally, a representative light-front wave function for the scalar deuteron at $E_B = 200$ MeV is displayed in Fig.~\ref{fig:deuteron-LFWF-EB200}.

\section{Summary and Outlook} \label{sec:summaries-and-outlooks}

In this work, we investigated the flavor asymmetry of the nucleon sea using the pion cloud model within the framework of a scalar EFT. By employing light-front wave functions solved non-perturbatively -- including Fock sectors with up to three pions -- we examined the influence of multi-pion effects on the nucleon's longitudinal momentum distributions. We observed robust numerical convergence at the two-pion level and have adopted the three-pion truncation as our final non-perturbative result. Our findings indicate that while both the leading-order perturbative and the non-perturbative approaches provide a reasonable description of the $\bar{d}-\bar{u}$ experimental data, the non-perturbative treatment significantly improves the description of the $\bar{d}/\bar{u}$ ratio towards the experimental value. Furthermore, the non-perturbative LMDs deviate substantially from their perturbative counterparts, underscoring the vital importance of multi-pion contributions to the proton's internal structure. 

We also explored the extension of this framework to describe the deuteron bound state. As a benchmark, we solved the two-body effective Hamiltonian derived via a similarity transformation. Our preliminary results suggest that incorporating multi-pion components is essential for a comprehensive understanding of flavor asymmetry in light nuclei. In the present calculations, dynamical pions have not yet been fully integrated into the nuclear bound state; however, the solution of the full four-body equation is currently in progress. The introduction of dynamical pions is expected to generate flavor asymmetry through a combination of pion cloud effects and boson-exchange mechanisms. Such a unified approach will be instrumental in clarifying the role of nuclear effects in flavor asymmetry and may help resolve the standing tensions between fixed-target experiments and recent LHC data.

\section*{Acknowledgements}
The authors wish to thank V. A. Karmanov for valuable discussions.
This work is supported in part by the NSFC under Grant No. 125B2111, and by the Chinese Academy of Sciences under Grant No. YSBR-101. 

\bibliographystyle{elsarticle-num}
\bibliography{ref}
\end{document}